\documentclass[a4paper,UKenglish,cleveref, autoref, thm-restate]{lipics-v2021}

\pdfoutput=1
\hideLIPIcs

\usepackage[T1]{fontenc}
\usepackage[utf8]{inputenc}
\usepackage{amssymb,mathtools}

\usepackage{xcolor}
\definecolor{keywordcolor}{rgb}{0.7, 0.1, 0.1}   
\definecolor{tacticcolor}{rgb}{0.0, 0.1, 0.6}    
\definecolor{commentcolor}{rgb}{0.4, 0.4, 0.4}   
\definecolor{symbolcolor}{rgb}{0.0, 0.1, 0.6}    
\definecolor{sortcolor}{rgb}{0.1, 0.5, 0.1}      
\definecolor{attributecolor}{rgb}{0.7, 0.1, 0.1} 

\usepackage{listings}

\DeclareRobustCommand*{\lean}{%
  \ifmmode
    \let\SavedBGroup\bgroup
    \def\bgroup{%
      \let\bgroup\SavedBGroup
      \hbox\bgroup
    }%
  \fi
  \lstinline
}

\DeclareUnicodeCharacter{22A8}{$\models$}
\DeclareUnicodeCharacter{02E2}{$^s$}
\DeclareUnicodeCharacter{02B7}{$^w$}
\DeclareUnicodeCharacter{22A4}{$\top$}

\DeclareMathOperator{\LTLnext}{\mathrm{X}}
\DeclareMathOperator{\LTLsnext}{\mathrm{X}^s}
\DeclareMathOperator{\LTLwnext}{\mathrm{X}^w}
\newcommand{\LTLuntil}{\mathrel{\mathcal{U}}}
\newcommand{\LTLrelease}{\mathrel{\mathcal{R}}}
\DeclareMathOperator{\LTLfinally}{\mathrm{F}}
\DeclareMathOperator{\LTLglobally}{\mathrm{G}}

\category{Short Paper}

\title{LeanLTL: A unifying framework for linear temporal logics in Lean (Short Paper)}

\author{Eric Vin}{University of California, Santa Cruz, USA \and \url{https://eric-vin.github.io/}}{evin@ucsc.edu}{https://orcid.org/0000-0002-3089-1129}{}

\author{Kyle A. Miller}{University of California, Santa Cruz, USA \and \url{https://kmill.github.io/}}{kymiller@ucsc.edu}{https://orcid.org/0000-0001-7400-5304}{}

\author{Daniel J. Fremont}{University of California, Santa Cruz, USA \and \url{https://people.ucsc.edu/~dfremont/}}{dfremont@ucsc.edu}{https://orcid.org/0000-0002-9992-9965}{}

\authorrunning{E.\ Vin, K.\ A.\ Miller, and D.\ J.\ Fremont}

\Copyright{Eric Vin, Kyle A.\ Miller, and Daniel J.\ Fremont}



\funding{This material is based upon work supported by the National Science Foundation under Award No. 2303564.}

\nolinenumbers

\EventEditors{John Q. Open and Joan R. Access}
\EventNoEds{2}
\EventLongTitle{42nd Conference on Very Important Topics (CVIT 2016)}
\EventShortTitle{CVIT 2016}
\EventAcronym{CVIT}
\EventYear{2016}
\EventDate{December 24--27, 2016}
\EventLocation{Little Whinging, United Kingdom}
\EventLogo{}
\SeriesVolume{42}
\ArticleNo{23}

\begin{document}

\maketitle

\begin{abstract}
We propose LeanLTL, a unifying framework for linear temporal logics in Lean 4. LeanLTL supports reasoning about traces that represent either infinite or finite linear time. The library allows traditional LTL syntax to be combined with arbitrary Lean expressions, making it straightforward to define properties involving numerical or other types. We prove that standard flavors of LTL can be embedded in our framework. The library also provides automation for reasoning about LeanLTL formulas in a way that facilitates using Lean's existing tactics. Finally, we provide examples illustrating the utility of the library in reasoning about systems that come from applications.
\end{abstract}

\section{Introduction}
\label{sec:introduction}
Linear temporal logic (LTL)~\cite{LTL_Pnueli_77, LTL_Piterman_18} has long been used in the verification community to specify the behaviors of systems over time.
LTL has achieved such longevity by striking a balance between expressivity --- for example being able to express convenient properties such as ``$\lean{CarAtDestination}$ will eventually be true forever'' and ``$\lean{GateOpen}$ is always eventually true'' --- and simplicity, being a decidable logic~\cite{LTL_Piterman_18} with practical tools for satisfiability checking, model checking, and synthesis (e.g.~\cite{PRISM_Kwiatkowska_11, BLACK_Geatti_21, SPIN_Holzmann_97, Strix_Meyer_18}).
LTL has been repeatedly used to great effect in real-world applications such as communication protocols~\cite{PRISMProtocols_Daws_04}, hardware synthesis~\cite{PSLHardwareSynthesis_Bloem_02}, security~\cite{PRISMSecurity_Norman_03}, and planning~\cite{PRISMControllersSynthesis_Feng_15}.

Applications of LTL often require variations on or extensions to the core logic.
For example, in runtime monitoring~\cite{LTLRV_Bauer_10} one needs to describe system behaviors over time periods of \emph{finite} extent, rather than the infinite traces used in LTL.
The logic LTLf~\cite{LTLf_Giacomo_13} accommodates this need; like LTL, it is decidable~\cite{LTLf_Giacomo_13} and efficiently solvable with existing tools~\cite{BLACK_Geatti_21}.

Basic LTL only supports propositional (boolean) variables, but, more generally, system specifications often involve predicates over integers, real numbers, arrays, and other data types.
For example, in LTL the property ``Eventually, $\lean{DistanceToCar} < 100$'' would be encoded as $\LTLfinally p$ with $p$ an atomic proposition defined outside of LTL itself to be true at exactly those timesteps where $\lean{DistanceToCar} < 100$.
The LTL encoding does not represent the relationship between the proposition $p$ and the underlying attribute of the system $\lean{DistanceToCar}$, meaning that some valid deductions cannot be carried out within LTL.
For example, if we know that $\lean{DistanceToCar}$ is either zero or decreases by at least 1 in each timestep of an infinite trace, then the property above would hold; but LTL cannot encode the conditions on $\lean{DistanceToCar}$ more precisely than as independent, opaque propositions.
LTL Modulo Theories (LTLMT)~\cite{LTLMT_Rodriguez_23} and LTLf Modulo Theories (LTLfMT)~\cite{LTLfMT_Geatti_16} address this shortcoming by replacing propositions with atoms from a given \emph{theory} in the vein of Satisfiability Modulo Theories (SMT)~\cite{SMT_Barrett_18}.
Commonly-used theories include linear integer arithmetic (LIA, a.k.a.~Presburger arithmetic), bitvectors (BV), and nonlinear real arithmetic (NRA).
When the underlying theory is decidable, there is a semi-decision procedure for LTLMT~\cite{LTLfMT_Geatti_16}, with certain theories supporting a full decision procedure~\cite{LTLfMTDecidability_Geatti_23}.
Returning to our previous example, we could represent and solve the implication ``($\lean{DistanceToCar}$ is $0$ or decreases by at least $1$) implies (Eventually, $\lean{DistanceToCar} < 100$)'' by encoding it as an LTLMT problem that utilizes the decidable theory of Linear Real Arithmetic~\cite{SMT_Barrett_18}.

Prior work on increasing the expressivity of linear temporal logics has largely attempted to stay within the realm of (semi-)decidability, to ensure that tools for the new logics can be fully automatic.
However, such approaches rule out the use of many useful undecidable theories, such as nonlinear real arithmetic with exponents or trigonometric functions (often needed for modeling cyber-physical systems).
Another limitation is that decision procedures for LTLMT are not available for all decidable theories~\cite{LTLfMTDecidability_Geatti_23}.
Finally, even in decidable cases, the complexity of the decision procedure can render realistic problems intractable in practice.
The above problems arise primarily in the context of \textit{automated} tools, but with interactive theorem provers one may permit undecidable theories, as users may provide proofs for what is not in the purview of decision procedures.

With this goal in mind, we propose LeanLTL\footnote{Available at: \url{https://github.com/UCSCFormalMethods/LeanLTL}. The exact version presented in this paper is tagged \texttt{vITP25}.}, a unified framework for reasoning about linear temporal properties of systems in Lean~4~\cite{Lean4}.
Its key features include:
\begin{itemize}
    \item Core types for modeling temporal properties across both infinite and finite traces, with support for arbitrary Lean expressions inside these formulas.
    Proofs using these types can make use of the full power of existing Lean tactics, as illustrated in \Cref{fig:teaser_example}.
    \item Convenient syntax for creating LeanLTL formulas with the appropriate semantics represented directly in Lean, with embedded Lean propositions.
    \item Basic automation and tools to simplify reasoning about LeanLTL formulas.
    \item Formalized proofs that LTL and LTLf can be embedded into LeanLTL.
\end{itemize}
In the rest of the paper, we describe these features and present real-world-inspired examples that illustrate the utility of LeanLTL.
Our library, examples, and proofs are included in the project repository.

\begin{figure}
    \centering
\begin{lstlisting}
example : ⊨ⁱ LLTL[((←n) = 5 ∧ G ((X (←n)) = (←n) ^ 2)) → G (5 ≤ (←n))] := by
    rw [TraceSet.sem_entail_inf_iff]
    rintro t hinf ⟨h1, h2⟩
    apply TraceSet.globally_induction <;> simp_all [push_ltl, hinf]
    intros; nlinarith\end{lstlisting}
    \caption{A proof in LeanLTL that for all infinite traces with a natural number variable $n$, the LTL-with-nonlinear-arithmetic formula $n=5 \land \LTLglobally((\LTLnext n) = n^2) \rightarrow \LTLglobally(n \ge 5)$ holds.}
    \label{fig:teaser_example}
\end{figure}

\vspace{0.25em}

\noindent \textbf{Related Work.}
Prior work on temporal logics in interactive theorem provers includes published work~\cite{TLPVS_Pnueli_03, ATTLCoq_Zanarini_12, MLTLIsabelle_Kosaia_25, CoqCTL_Doczkal_16} and open-source repositories~\cite{GitLTL_UnitB_24, GitLTL_Murphy_24, GitLTL_Oswald_24, GitLTL_Galois_25}.
To our knowledge, our work is the first framework in Lean to support discrete-time linear temporal logics over both finite and infinite traces while allowing full utilization of arbitrary underlying theories.
The most similar work to ours is the TLPVS system~\cite{TLPVS_Pnueli_03}, a library for reasoning about LTL properties in PVS~\cite{PVS_Owre_92} with support for theories, but it does not consider finite traces.

\section{Background}
\label{sec:background}
In this section we summarize the syntax and semantics of LTL and some of its extensions.

A full definition of the syntax and semantics of LTL can be found in \cite{LTL_Piterman_18}. LTL includes the standard propositional operators NOT ($\lnot$), AND ($\land$), and OR ($\lor$), and the constant True ($\top$), in addition to several \textit{temporal operators}:
\begin{itemize}
    \item Next: $\LTLnext p$ (or $\bigcirc p$) is true if and only if $p$ is true in the next timestep.
    \item Until: $p \LTLuntil q$ is true iff $p$ is true at least until $q$ is true, and $q$ must eventually be true.
    \item Release: $p \LTLrelease q$ is true iff $p$ is true up until and including the point where $q$ is true. If $q$ never becomes true, $p$ must be true forever.
    \item Finally: $\LTLfinally p$ (or $\lozenge p$) is true iff $p$ is true in some future timestep, equivalent to $\top \LTLuntil p$.
    \item Globally: $\LTLglobally p$ (or $\square p$) is true iff $p$ is true in all future timesteps, equivalent to $\lnot \LTLfinally \lnot p$.
\end{itemize}

LTLf, whose full syntax and semantics can be found in \cite{LTLf_Giacomo_13}, modifies the semantics of $\LTLnext$ to account for the possibility that the next value does not exist due to being at the end of a finite trace.
The two resulting operators are \emph{strong next}, with $\LTLsnext p$ ($\bigcirc p$ in \cite{LTLf_Giacomo_13}) evaluating to false at the end of a finite trace, and \emph{weak next}, with $\LTLwnext p$ (originally $\ooalign{\hss$\sim$\hss\cr$\bigcirc$} p$) evaluating to true at the end of a finite trace; it is equivalent to $(\LTLsnext p)\lor\neg\LTLsnext\top$.

LTLfMT expands these semantics further by replacing propositions with atomic formulas from an underlying theory, allowing $\LTLsnext$ and $\LTLwnext$ to be applied to terms inside those formulas.
A full definition of its syntax and semantics can be found in \cite{LTLfMT_Geatti_16}.

\section{Library Details and Implementation}
\label{sec:library_details}

In this section we describe the key components of LeanLTL: types for traces and formulas, a macro enabling the use of traditional LTL syntax, and automation for proofs involving LTL.

\subsection{Core Types}
\label{sec:core-types}

\textbf{Traces.}
To model properties across linear time, we use sequences of values called \emph{traces}.
Often LTL formulas are about traces over a \emph{state} type, a structure containing fields for all the time-varying propositions and values under consideration, and we model these time-varying variables as projections of a trace of states (see \Cref{sec:evaluation} for an example).

Concretely, we define traces over a type $\sigma$ in the following way:
\begin{lstlisting}
structure Trace (σ : Type*) where
  toFun? : ℕ → Option σ
  length : ℕ∞
  nempty : 0 < length
  defined : ∀ i : ℕ, i < length ↔ (toFun? i).isSome\end{lstlisting}
This models both finite and infinite traces by using an option-valued sequence \lean{toFun?}, which has the property that if it ever becomes \lean{none} it remains \lean{none} for all time (enforced by the \lean{defined} proof).
The extended natural number \lean{length} records the length of the trace for convenience; all of our \lean{Trace} constructions yield concrete values for the \lean{length} field.

In order to ensure propositions always have a defined truth value, we require traces to be nonempty, as in LTLf.
Reasoning about possibly-undefined truth values only appears in conjunction with the strong/weak next operators.
We remark that if LTLf had notions of ``strong/weak evaluation'' of propositional variables, one could relax the nonempty constraint and use the LTL next operator semantics. 

We model moving forward in time by shift operators on traces, which drop some number of terms from the beginning of the sequence.
These operators require proof that the shift results in a nonempty trace.
We also provide a number of operations to support constructions for the \lean{TraceSet} and \lean{TraceFun} types, as we will see below.\\

\noindent \textbf{Trace Sets.}
We take a semantic view of LTL formulas, representing a formula by its set of satisfying traces, which we call a \textit{trace set}.
\begin{lstlisting}
structure TraceSet (σ : Type*) where
  sat : Trace σ → Prop

notation t " ⊨ " p => TraceSet.sat p t\end{lstlisting}
This type is equivalent to \lean{Set (Trace σ)}, using \lean{Set} from Lean's Mathlib, but a custom structure gives us some advantages: (1) we can use \lean{t ⊨ p} notation for satisfaction without it being pretty-printed as \lean{t ∈ p}, (2) we can ensure it is never confused for \lean{Set} in any Lean automation, and (3) we can make heavy use of \emph{generalized field notation} by putting declarations in the \lean{TraceSet} namespace:
for example, we can write \lean{p.implies (q.and r)} for \lean{TraceSet.implies p (TraceSet.and q r)}.
As we will describe in \Cref{sec:macro}, we can further simplify such notation with a macro enabling the use of Lean's own logical connectives.\\

\noindent \textbf{Trace Functions.}
To enable encoding logical formulas embedding Lean propositions which refer to the current trace, we introduce \emph{trace functions}, which are simply functions from traces on a given domain to an option type:
\begin{lstlisting}
structure TraceFun (σ α : Type*) where
  eval : Trace σ → Option α
\end{lstlisting}
The \lean{none} value is intended to indicate exceptional behavior, such as needing to access a value past the end of a finite trace.
Every \lean{TraceSet σ} can be cast as a \lean{TraceFun σ Prop}, but going in the reverse direction needs a choice of whether to interpret \lean{none} as \lean{True} or \lean{False}.

The main interface between trace functions and trace sets are the \lean{TraceFun.sget} and \lean{TraceFun.wget} functions (\emph{strong get} and \emph{weak get}), which bind the value of a trace function \lean{f : TraceFun σ α} to a variable that can then be used inside a trace set, where the first evaluates to \lean{False} on \lean{none} and the second to \lean{True}.
For example, with $\alpha=\mathbb{N}$, we may write
\begin{lstlisting}
t ⊨ TraceFun.sget f (fun x => TraceSet.const (x < 10))\end{lstlisting}
and, if the value of \lean{f} exists, it is compared to $10$, and otherwise the proposition evaluates to false as this is a strong get.
We will see a convenient notation for this in \Cref{sec:macro}.

The main trace functions are \lean{TraceFun.of}, for projecting a value from the current state, and \lean{TraceFun.shift}, for precomposing trace functions with a trace shift operator, which models having the trace function look forward in time.

\subsection{LeanLTL Macro Syntax}
\label{sec:macro}
To aid in writing LeanLTL formulas, we offer an \lean{LLTL[...]} macro that reinterprets Lean's logical connectives as LeanLTL formulas, giving a seamless embedding of Lean's logic while exposing the ability to use trace functions embedded in Lean expressions.
The macro can be compared to idiom brackets in~\cite{McBride2008}, and the embedded trace functions can be compared to the \verb|!x| notation in Idris~\cite{Idris} and the \lean{(← x)} notation in Lean for embedded monad expressions.

The macro has two main components.
First, we have a mechanism to install interpretations of host connectives as LTL connectives.
For example,
\begin{lstlisting}
declare_lltl_notation p q : p → q => TraceSet.imp p q\end{lstlisting}
causes \lean{LLTL[x → y]} to expand as \lean{TraceSet.imp x y}, while also creating a pretty printer making \lean{TraceSet.imp} display using the \lean{LLTL} macro.
This macro supports Lean's editor integrations, and hovering over \lean{→} in the Infoview gives type information, as if it were written as \lean{TraceSet.imp}.
Second, all Lean expressions that do not have a registered interpretation are assumed to be Lean propositions, possibly with embedded \lean{TraceFun} expressions.
Each \lean{(←ˢ f)} (synonym: \lean{(← f)}) and \lean{(←ʷ f)} is lifted out using the \lean{TraceFun.sget} and \lean{TraceFun.wget} functions mentioned in \Cref{sec:core-types}.
For example,
\lstinline{t ⊨ LLTL[G ((←ˢ f) < 10)]}
expands to
\begin{lstlisting}
t ⊨ TraceSet.globally (TraceFun.sget f fun x => TraceSet.const (x < 10))\end{lstlisting}
The left arrows should be compared to Lean's monad arrow expressions, which in monadic contexts are similarly lifted out of expressions using monadic binds.
If there are multiple strong gets and weak gets, we use the convention that strong gets are bound first, which reliably causes the surrounding LTL formula to evaluate to false at the end of a trace, no matter the order in which the arrow expressions appear inside the embedded Lean formula.

The macro expansion of \lean{LLTL[...]} notation is carried out using recursively applied rules.
First, the \lean{declare_lltl_notation} macros are applied; for example, \lean{LLTL[G ((←ˢ f) < 10)]} is first expanded to \lean{TraceSet.globally LLTL[(←ˢ f) < 10]}.
Second, \lean{X} notation is pushed into arrow notations, so for example \lean{LLTL[X ((← a) < (← b))]} becomes \lean{LLTL[(← X a) < (← X b))]}, which implements $\LTLnext$ for compound values without needing to use \lean{TraceFun} analogues of Lean operators.
Third, arrow notations are lifted out, for example \lean{LLTL[(←ˢ f) < 10]} becomes \lean{TraceFun.sget LLTLV[f] fun x => LLTL[x < 10]}, where \lean{LLTLV[...]} is a similar macro for \lean{TraceFun}.
Finally, whatever remains is coerced to a \lean{TraceSet}.
Pure propositions use \lean{TraceSet.const}, and we also coerce predicates \lean{σ → Prop} to \lean{TraceSet σ} using \lean{TraceSet.of}, which enables the direct use of structure projections in the formulas in \Cref{fig:tl_example_code}.
The \lean{LLTLV[...]} macro expands similarly and implements similar coercions.

\subsection{Automation}
\label{sec:automation}
We provide preliminary automation for reasoning about LeanLTL formulas, with the intention of expanding on this as we continue development.
Our initial automation is centered on \emph{simp sets}, curated sets of theorems that can be provided to the \lean{simp} tactic. The primary simp set is \lean{push_ltl}, which ``pushes'' the LTL ``satisfies'' operation as deep as possible into an expression, translating LTL operations into their first-order logic semantics.

This simp set can often reduce problems to a form that other tactics such as \lean{linarith} or \lean{omega} can finish, providing a halfway point between the semi/full decision procedures supported by solvers of LTLfMT.
\Cref{fig:teaser_example} is an example, where after applying the \lean{TraceSet.globally_induction} principle (provided by LeanLTL) for proving formulas of the form $\LTLglobally p$ by induction on time,
the \lean{push_ltl} simp set discharges the base case and leaves the following goal:
\begin{lstlisting}
h1 : n (t.toFun 0 ⋯) = 5
h2 : ∀ (n : ℕ), n (t.toFun (1 + n) ⋯) = n (t.toFun n ⋯) ^ 2
⊢ ∀ (n : ℕ), 5 ≤ n (t.toFun n ⋯) → 5 ≤ n (t.toFun n ⋯) ^ 2\end{lstlisting}
These nonlinear inequalities are then solved using the \lean{nlinarith} tactic.

In future work, we aim to integrate LTL and LTLMT decision procedures directly into our library.

\section{Logic Embeddings and Examples}
\label{sec:evaluation}
In this section, we illustrate the expressivity of LeanLTL and its ability to help prove useful properties about real-world systems.
For the former, we provide Lean proofs that LTL and LTLf can be embedded into LeanLTL.
The embedding proof first provides an inductive syntax for each logic and an evaluation function giving its semantics.
We then define a function translating the logic to LeanLTL.
Finally, we show that the evaluation function applied to the original syntax is equisatisfiable with our LeanLTL translation and the LeanLTL semantics. An abbreviated version of the LTL embedding proof can be found in Figure \ref{fig:embedding_code}, and the complete Lean proofs are included in the project repository.

\begin{figure}[t]
    \centering
    \begin{lstlisting}
def Var (σ : Type*) := σ -> Prop

structure Trace (σ : Type*) where
  trace : LeanLTL.Trace σ
  infinite : trace.Infinite

inductive Formula (σ : Type*) where
  | var (v : (Var σ))
  | not (f : Formula σ)
  | or (f₁ f₂ : Formula σ)
  | next (f : Formula σ)
  | until (f₁ f₂ : Formula σ)

def sat {σ : Type*} (t : Trace σ) (f : Formula σ) : Prop := by ...
def toLeanLTL {σ : Type*} (f : Formula σ) : LeanLTL.TraceSet σ := ...

theorem equisat {σ : Type*} (f : Formula σ) (t : Trace σ) : 
    sat t f ↔ (t.trace ⊨ toLeanLTL f) := by ...
\end{lstlisting}
    \caption{A high-level skeleton of the LTL embedding into \lean{TraceSet}.}
    \label{fig:embedding_code}
\end{figure}

To illustrate the application of LeanLTL to non-trivial real-world scenarios, we provide an abbreviated example applying the library to a system consisting of two traffic lights at an intersection, with which we demonstrate how LeanLTL can be used to model a system, its environment, and its specifications, and prove that a specification is satisfied for any given trace. The full example can be found in the supplemental materials.

To begin, we declare the state of the system at each timestep, shown in the first part of \Cref{fig:tl_example_code}.
The state consists of whether each of the two lights is green, as well as the numbers of cars arriving, leaving, or waiting at each light.

Next we define properties that we assume about the system.
In our example these properties are assumed to be derived from external specifications, but they could also include a Lean function representing a part of the system used directly in a LeanLTL formula.
The second part of \Cref{fig:tl_example_code} shows a selection of these properties: for example, \lean{TL1ToTL2Green} states that when light 1 is green and has no cars waiting, in the next timestep it turns red and light 2 turns green.
Note the use of arithmetic here and in \lean{TL1QueueNext}.

In the last part of the figure, we state the properties that we wish to prove our system satisfies:  \lean{G_OneLightGreen} states that exactly one of the traffic lights is green at any given time, and \lean{G_F_Green} states that each light will turn green infinitely often.

Finally, we prove that our desired properties hold for all traces satisfying the assumed properties.
Complete proofs can be found in the project repository.

\begin{figure}[tb]
    \centering
    \begin{lstlisting}
-- State structure, defining the data available at each timestep
structure ExState where
  (TL1Green TL2Green : Prop)
  (TL1Arrives TL1Departs TL1Queue : ℕ)
  (TL2Arrives TL2Departs TL2Queue : ℕ)
open ExState
-- Assumed properties, describing the behavior of the system
abbrev TL1StartGreen    := LLTL[TL1Green]
abbrev TL1ToTL2Green    :=
  LLTL[G ((TL1Green ∧ (← TL1Queue) = 0) → (Xˢ (¬TL1Green ∧ TL2Green)))]
abbrev TL1StayGreen     :=
  LLTL[G ((TL1Green ∧ (← TL1Queue) ≠ 0) → (Xˢ (TL1Green ∧ ¬ TL2Green)))]
abbrev TL1GreenDeparts  := LLTL[G (TL1Green → (← TL1Departs) = max_departs)]
abbrev TL1RedDeparts    := LLTL[G (¬TL1Green → (← TL1Departs) = 0)]
abbrev TL1ArrivesBounds :=
  LLTL[G (0 ≤ (← TL1Arrives) ∧ (← TL1Arrives) ≤ max_arrives)]
abbrev TL1QueueNext     :=
  LLTL[G ((X (← TL1Queue)) = (← TL1Queue) + (← TL1Arrives) - (← TL1Departs))]
-- Desired properties, which we prove
abbrev G_OneLightGreen  := LLTL[G (TL1Green ↔ ¬TL2Green)]
abbrev G_F_Green        := LLTL[(G (F TL1Green)) ∧ (G (F TL2Green))]
\end{lstlisting}
    \caption{The state structure and some assumed/desired properties in the traffic light example.}
    \label{fig:tl_example_code}
\end{figure}

\section{Conclusion}
In this paper we presented our initial prototype of LeanLTL, a unifying framework for linear temporal logics in Lean.
We described the core features of the library and illustrated how they can be used to encode and prove temporal properties of systems in Lean.
In future work we plan to develop further automation, including integrations with decision procedures, as well as embedding proofs for LTLMT and LTLfMT.
We also plan to consider adding support for other LTL variants, including past-time and bounded-time operators.



\begin{thebibliography}{10}

\bibitem{SMT_Barrett_18}
Clark Barrett and Cesare Tinelli.
\newblock Satisfiability {Modulo} {Theories}.
\newblock In Edmund~M. Clarke, Thomas~A. Henzinger, Helmut Veith, and Roderick Bloem, editors, {\em Handbook of {Model} {Checking}}, pages 305--343. Springer International Publishing, Cham, 2018.
\newblock URL: \url{https://doi.org/10.1007/978-3-319-10575-8_11}.

\bibitem{LTLRV_Bauer_10}
Andreas Bauer, Martin Leucker, and Christian Schallhart.
\newblock Comparing {LTL} {Semantics} for {Runtime} {Verification}.
\newblock {\em Journal of Logic and Computation}, 20(3):651--674, June 2010.
\newblock \href {https://doi.org/10.1093/logcom/exn075} {\path{doi:10.1093/logcom/exn075}}.

\bibitem{PSLHardwareSynthesis_Bloem_02}
Roderick Bloem, Stefan Galler, Barbara Jobstmann, Nir Piterman, Amir Pnueli, and Martin Weiglhofer.
\newblock Specify, {Compile}, {Run}: {Hardware} from {PSL}.
\newblock {\em Electronic Notes in Theoretical Computer Science}, 190(4):3--16, November 2007.
\newblock URL: \url{https://www.sciencedirect.com/science/article/pii/S157106610700583X}, \href {https://doi.org/10.1016/j.entcs.2007.09.004} {\path{doi:10.1016/j.entcs.2007.09.004}}.

\bibitem{Idris}
Edwin Brady.
\newblock Idris, a general-purpose dependently typed programming language: Design and implementation.
\newblock {\em Journal of Functional Programming}, 23:552--593, 9 2013.
\newblock URL: \url{https://journals.cambridge.org/article_S095679681300018X}, \href {https://doi.org/10.1017/S095679681300018X} {\path{doi:10.1017/S095679681300018X}}.

\bibitem{PRISMProtocols_Daws_04}
Conrado Daws, Marta Kwiatkowska, and Gethin Norman.
\newblock Automatic verification of the {IEEE} 1394 root contention protocol with {KRONOS} and {PRISM}.
\newblock {\em International Journal on Software Tools for Technology Transfer}, 5(2):221--236, March 2004.
\newblock \href {https://doi.org/10.1007/s10009-003-0118-5} {\path{doi:10.1007/s10009-003-0118-5}}.

\bibitem{LTLf_Giacomo_13}
Giuseppe De~Giacomo and Moshe~Y. Vardi.
\newblock Linear temporal logic and linear dynamic logic on finite traces.
\newblock In {\em Proceedings of the {Twenty}-{Third} international joint conference on {Artificial} {Intelligence}}, {IJCAI} '13, pages 854--860, Beijing, China, August 2013. AAAI Press.

\bibitem{CoqCTL_Doczkal_16}
Christian Doczkal and Gert Smolka.
\newblock Completeness and {Decidability} {Results} for {CTL} in {Constructive} {Type} {Theory}.
\newblock {\em Journal of Automated Reasoning}, 56(3):343--365, March 2016.
\newblock \href {https://doi.org/10.1007/s10817-016-9361-9} {\path{doi:10.1007/s10817-016-9361-9}}.

\bibitem{PRISMControllersSynthesis_Feng_15}
Lu~Feng, Clemens Wiltsche, Laura Humphrey, and Ufuk Topcu.
\newblock Controller synthesis for autonomous systems interacting with human operators.
\newblock In {\em Proceedings of the {ACM}/{IEEE} {Sixth} {International} {Conference} on {Cyber}-{Physical} {Systems}}, {ICCPS} '15, pages 70--79, New York, NY, USA, April 2015. Association for Computing Machinery.
\newblock URL: \url{https://dl.acm.org/doi/10.1145/2735960.2735973}, \href {https://doi.org/10.1145/2735960.2735973} {\path{doi:10.1145/2735960.2735973}}.

\bibitem{GitLTL_Galois_25}
{Galois, Inc.}
\newblock lean-protocol-support/galois/temporal at master · {GaloisInc}/lean-protocol-support.
\newblock URL: \url{https://github.com/GaloisInc/lean-protocol-support/tree/master/galois/temporal}.

\bibitem{LTLfMT_Geatti_16}
Luca Geatti, Alessandro Gianola, and Nicola Gigante.
\newblock Linear {Temporal} {Logic} {Modulo} {Theories} over {Finite} {Traces}.
\newblock In {\em Proceedings of the {Thirty}-{First} {International} {Joint} {Conference} on {Artificial} {Intelligence}}, volume~3, pages 2641--2647, July 2022.
\newblock URL: \url{https://www.ijcai.org/proceedings/2022/366}, \href {https://doi.org/10.24963/ijcai.2022/366} {\path{doi:10.24963/ijcai.2022/366}}.

\bibitem{LTLfMTDecidability_Geatti_23}
Luca Geatti, Alessandro Gianola, Nicola Gigante, and Sarah Winkler.
\newblock Decidable {Fragments} of {LTLf} {Modulo} {Theories}.
\newblock In {\em {ECAI} 2023}, pages 811--818. IOS Press, 2023.
\newblock URL: \url{https://ebooks.iospress.nl/doi/10.3233/FAIA230348}.

\bibitem{BLACK_Geatti_21}
Luca Geatti, Nicola Gigante, and Angelo Montanari.
\newblock {BLACK}: {A} {Fast}, {Flexible} and {Reliable} {LTL} {Satisfiability} {Checker}.
\newblock {\em CEUR Workshop Proceedings}, September 2021.

\bibitem{SPIN_Holzmann_97}
G.J. Holzmann.
\newblock The model checker {SPIN}.
\newblock {\em IEEE Transactions on Software Engineering}, 23(5):279--295, May 1997.
\newblock URL: \url{https://ieeexplore.ieee.org/document/588521}, \href {https://doi.org/10.1109/32.588521} {\path{doi:10.1109/32.588521}}.

\bibitem{GitLTL_UnitB_24}
Simon Hudon.
\newblock unitb/temporal-logic, November 2024.
\newblock URL: \url{https://github.com/unitb/temporal-logic}.

\bibitem{MLTLIsabelle_Kosaia_25}
Katherine Kosaian, Zili Wang, Elizabeth Sloan, and Kristin Rozier.
\newblock Formalizing {MLTL} {Formula} {Progression} in {Isabelle}/{HOL}, February 2025.
\newblock URL: \url{http://arxiv.org/abs/2410.03465}, \href {https://doi.org/10.48550/arXiv.2410.03465} {\path{doi:10.48550/arXiv.2410.03465}}.

\bibitem{PRISM_Kwiatkowska_11}
Marta Kwiatkowska, Gethin Norman, and David Parker.
\newblock {PRISM} 4.0: {Verification} of {Probabilistic} {Real}-{Time} {Systems}.
\newblock In Ganesh Gopalakrishnan and Shaz Qadeer, editors, {\em Computer {Aided} {Verification}}, pages 585--591, Berlin, Heidelberg, 2011. Springer.
\newblock \href {https://doi.org/10.1007/978-3-642-22110-1_47} {\path{doi:10.1007/978-3-642-22110-1_47}}.

\bibitem{McBride2008}
Conor McBride and Ross Paterson.
\newblock Applicative programming with effects.
\newblock {\em Journal of Functional Programming}, 18(1):1–13, 2008.
\newblock \href {https://doi.org/10.1017/S0956796807006326} {\path{doi:10.1017/S0956796807006326}}.

\bibitem{Strix_Meyer_18}
Philipp~J. Meyer, Salomon Sickert, and Michael Luttenberger.
\newblock Strix: {Explicit} {Reactive} {Synthesis} {Strikes} {Back}!
\newblock In Hana Chockler and Georg Weissenbacher, editors, {\em Computer {Aided} {Verification}}, pages 578--586, Cham, 2018. Springer International Publishing.
\newblock \href {https://doi.org/10.1007/978-3-319-96145-3_31} {\path{doi:10.1007/978-3-319-96145-3_31}}.

\bibitem{Lean4}
Leonardo~de Moura and Sebastian Ullrich.
\newblock The {L}ean 4 theorem prover and programming language.
\newblock In Andr{\'e} Platzer and Geoff Sutcliffe, editors, {\em Automated Deduction -- CADE 28}, pages 625--635, Cham, 2021. Springer International Publishing.

\bibitem{GitLTL_Murphy_24}
Logan Murphy.
\newblock loganrjmurphy/lean-temporal, November 2024.
\newblock URL: \url{https://github.com/loganrjmurphy/lean-temporal}.

\bibitem{PRISMSecurity_Norman_03}
Gethin Norman and Vitaly Shmatikov.
\newblock Analysis of {Probabilistic} {Contract} {Signing}.
\newblock In Ali~E. Abdallah, Peter Ryan, and Steve Schneider, editors, {\em Formal {Aspects} of {Security}}, pages 81--96, Berlin, Heidelberg, 2003. Springer.
\newblock \href {https://doi.org/10.1007/978-3-540-40981-6_9} {\path{doi:10.1007/978-3-540-40981-6_9}}.

\bibitem{GitLTL_Oswald_24}
James Oswald.
\newblock James-{Oswald}/linear-temporal-logic, November 2024.
\newblock URL: \url{https://github.com/James-Oswald/linear-temporal-logic}.

\bibitem{PVS_Owre_92}
Sam Owre, John~M. Rushby, and Natarajan Shankar.
\newblock {PVS}: {A} {Prototype} {Verification} {System}.
\newblock In {\em Proceedings of the 11th {International} {Conference} on {Automated} {Deduction}: {Automated} {Deduction}}, {CADE}-11, pages 748--752, Berlin, Heidelberg, June 1992. Springer-Verlag.

\bibitem{LTL_Piterman_18}
Nir Piterman and Amir Pnueli.
\newblock Temporal {Logic} and {Fair} {Discrete} {Systems}.
\newblock In Edmund~M. Clarke, Thomas~A. Henzinger, Helmut Veith, and Roderick Bloem, editors, {\em Handbook of {Model} {Checking}}, pages 27--73. Springer International Publishing, Cham, 2018.
\newblock URL: \url{https://doi.org/10.1007/978-3-319-10575-8_2}.

\bibitem{LTL_Pnueli_77}
Amir Pnueli.
\newblock The temporal logic of programs.
\newblock In {\em 18th {Annual} {Symposium} on {Foundations} of {Computer} {Science} (sfcs 1977)}, pages 46--57, October 1977.
\newblock URL: \url{https://ieeexplore.ieee.org/document/4567924}, \href {https://doi.org/10.1109/SFCS.1977.32} {\path{doi:10.1109/SFCS.1977.32}}.

\bibitem{TLPVS_Pnueli_03}
Amir Pnueli and Tamarah Arons.
\newblock {TLPVS}: {A} {PVS}-{Based} ltl {Verification} {System}.
\newblock In Nachum Dershowitz, editor, {\em Verification: {Theory} and {Practice}: {Essays} {Dedicated} to {Zohar} {Manna} on the {Occasion} of {His} 64th {Birthday}}, pages 598--625. Springer, Berlin, Heidelberg, 2003.
\newblock URL: \url{https://doi.org/10.1007/978-3-540-39910-0_26}.

\bibitem{LTLMT_Rodriguez_23}
Andoni Rodríguez and César Sánchez.
\newblock Boolean {Abstractions} for {Realizability} {Modulo} {Theories}.
\newblock In Constantin Enea and Akash Lal, editors, {\em Computer {Aided} {Verification}}, pages 305--328, Cham, 2023. Springer Nature Switzerland.
\newblock \href {https://doi.org/10.1007/978-3-031-37709-9_15} {\path{doi:10.1007/978-3-031-37709-9_15}}.

\bibitem{ATTLCoq_Zanarini_12}
Dante Zanarini, Carlos Luna, and Luis Sierra.
\newblock Alternating-{Time} {Temporal} {Logic} in the {Calculus} of ({Co}){Inductive} {Constructions}.
\newblock In Rohit Gheyi and David Naumann, editors, {\em Formal {Methods}: {Foundations} and {Applications}}, pages 210--225, Berlin, Heidelberg, 2012. Springer.
\newblock \href {https://doi.org/10.1007/978-3-642-33296-8_16} {\path{doi:10.1007/978-3-642-33296-8_16}}.

\end{thebibliography}
\end{document}